\documentclass{aastex}
\usepackage{emulateapj5}
\usepackage{apjfonts}
\newcommand{\um}{\mbox{\,$\mu$m}}
\shorttitle{First Detection of Millimeter Dust Emission from Brown Dwarf Disks}
\shortauthors{R. Klein et al.}
\begin{document}

\title{First Detection of Millimeter Dust Emission from Brown Dwarf Disks}

\author{R. Klein} \affil{Astrophysical Institute and University
  Observatory (AIU) Jena, Schillerg\"a{\ss}chen 2-3, D--07745
  Jena, Germany}\email{rklein@astro.uni-jena.de}
\author{D. Apai, I. Pascucci, Th.  Henning} \affil{Max Planck Institute for
  Astronomy, K\"onigstuhl 17, D--69117 Heidelberg, Germany}
\email{apai@mpia-hd.mpg.de;pascucci@mpia-hd.mpg.de;henning@mpia-hd.mpg.de}
\and
\author{L. B. F. M. Waters}\affil{Astronomical Institute ``Anton Pannekoek'',
  University of Amsterdam, Kruislaan 403, NL-1098 SJ Amsterdam,
  Netherlands\\ Institute of Astronomy, Katholieke Universiteit Leuven,
  Celestijnenlaan 200B, B3001 Heverlee, Belgie}
\email{rensw@science.uva.nl}

\begin{abstract}
  We report results from the first deep millimeter continuum survey
  targeting Brown Dwarfs (BDs). The survey led to the first detection
  of cold dust in the disks around two young BDs (CFHT-BD-Tau 4 and
  IC\,348\,613), with deep JCMT and IRAM observations reaching flux
  levels of a few mJy.  The dust masses are estimated to be a few
  Earth masses assuming the same dust opacities as usually applied to
  T\,Tauri stars.
\end{abstract}


\section{Introduction}

Several near-infrared surveys
\citep{Muench01,Oliveira02,Liu03,Jayawardhana03} and mid-infrared
measurements \citep{Comeron00,Persi00,Testi02,Natta02,Apai02} indicate
the presence of disks around Brown Dwarfs (BDs). In contrast to
infrared emission, submillimeter and millimeter emission is certainly
always optically thin and is an excellent measure for the total dust
mass. 

An early attempt to detect millimetre continuum emission also from
very low-mass stars and suspected BDs was done by \citet{Andre94}.
Probably the most sensitive observation of this kind was carried out
by \citet{Carpenter02} using the OVRO interferometer, however at a
relatively long wavelength.  We carried out the first successful
search for dust continuum emission associated with confirmed BDs,
using the bolometer arrays SCUBA at the JCMT and MAMBO at the IRAM
30-m telescope. 

The survey led to the detection of circumstellar dust around the two
young BDs CFHT-BD-Tau 4 and IC\,348\,613, which have ages below 10
Myrs.  In the case of field BDs we obtained upper mass limits of a few
Moon masses of dust. For BDs in the Pleiades the mass limits are less
strict and range 4 and 7 Earth masses. We should note that the data
presented for CFHT-BD-Tau 4 together with other ground-based and ISO
data allowed the first detailed discussion of a complete spectral
energy distribution of a BD, ranging from optical to millimeter
wavelengths \citep{Pascucci03}.

The detection of these amounts of circumstellar material around two
young BDs makes the formation of planets or even planetary systems
around BDs a possibility. Therefore, search strategies for planets
should include BDs. In the case of imaging surveys, they might even be
the best targets.

In addition, the detection of significant amount of circumstellar 
dust carries important information about the formation processes of BDs.
These detections, together with the discovery of quite a number of BD binaries 
\citep{Bouy03,Burgasser03}, do not support fragmentation of circumstellar
disks as the general process for BD formation. Other
formation scenarios for BDs \citep{Bate03, Reipurth01,Watkins98}
include ejection from multiple systems and
erosion of star-forming cloudlets by stellar winds and UV
radiation from massive stars. Disks will certainly have different
structures depending on the formation mechanism.  However,
statistics is still poor and information about the disk structure
from interferometric observations are needed before one can put
more definite observational constraints on BD formation scenarios.

\section{Observations}

To search for circumstellar material around BDs, we selected
relatively young BDs with an age of a few Myrs because these objects
should have the highest probability to be associated with disk
material.  The nine selected objects are located in Taurus, the
$\sigma$ Orionis cluster, IC 348, and the Upper Scorpius OB
association \citep{Martin01,Bejar01,Najita00,Luhman99,Ardila00}.  For
the first three regions, additional selection criteria were the
previous detection of H$\alpha$ emission \citep{Martin01}, the
presence of X-ray emission \citep{Mokler02,Preibisch01}, and the
requirement that the objects should be as isolated as possible in
order to avoid confusion during the observations with single-dish
telescopes.

The three BDs in the Taurus star-forming region are among the
youngest BDs of our  target list. They have ages of about 1\,Myr
\citep{Martin01}. The object CFHT-BD-Tau 4 shows the highest
H$\alpha$ emission among the Taurus BDs \citep{Martin01}, emits
X-ray radiation \citep{Mokler02} and shows mid-infrared excess
emission \citep{Pascucci03}.

The second group, we selected, are objects in the Pleiades which have
a distance of 116~pc and an age of 0.12 Gyr. The objects are
considerably older than the first group of BDs. In addition, we
searched for dust emission from very nearby field BDs. On the average,
these objects are older than the first group, too, but the actual age
determination contains large uncertainties.

The  targets with their ages and distances are compiled in Table
\ref{tab:obs}.

The observations were carried out at the James Clerk Maxwell
Telescope (JCMT) on Mauna Kea, Hawai'i, and at the IRAM 30\,m
telescope, Pico Veleta, Spain. At the JCMT we used the 37-channel
array camera SCUBA \citep{scuba} at a wavelength of 850\um{}. At
the IRAM 30-m telescope, we performed the observations at 1.3\,mm
with the 37-channel  array camera MAMBO \citep{Bolos}. These
wavelengths are a good tradeoff between the expected flux and the
opacity of the atmosphere.

The  observations were performed between March 2002 and January
2003. The chosen observing mode is photometry with the central
bolometer of SCUBA and MAMBO, respectively. The background 
(telluric and astronomical) subtraction is achieved by chopping 
with the secondary mirror and nodding the telescope. Its level has 
been estimated from the inner ring consisting of six bolometer
pixels. 

The measured flux densities and derived upper limits are listed in
Table~\ref{tab:res}.  For many sources, we can only derive upper
limits since the signal-to-noise-ratio (SNR) is less than three.
The upper limits are given as 3$\sigma$ values, where $\sigma$ is
the standard deviation of the statistical error.

\section{Results}
We detected millimeter continuum emission associated with two
BDs. Both, BD\,Tau\,4 and IC\,348\,613, were detected at two
wavelengths, i.e. with the JCMT at 850{\um} and with the IRAM
30m telescope at 1.3\,mm. These detections can be used to estimate the
amount of circumstellar matter around these BDs for the first time.

\citet{Carpenter02} however using the OVRO interferometer to map IC
348 did not detect IC\,348\,613. The reason is that he mapped at a
wavelength of 3\,mm.  If we extrapolate the measured flux densities to
a wavelength of 3\,mm, the flux density falls just below the detection
limit of the OVRO map.

\subsection{Dust Masses}

The measured flux density $F_\nu$ at submillimeter/millimeter
wavelengths is certainly optically thin thermal emission by cold dust
heated by the BD. The dust mass $M$ can be derived using the formula
\[M=\frac{F_\nu D^2}{B_\nu(T,\lambda)\kappa_\nu(\lambda)},\]
where $F_\nu$ stands for the flux density, $B_\nu$ for the Planck
function, $T$ for the dust temperature and $\kappa_\nu$ for the mass
absorption coefficient. The quantity $D$ is the distance to the
object.  Likewise upper limits for the flux density are translated
into upper limits for the amount of circumstellar matter.

Among the above quantities, the distances are well established from
trigonometric parallaxes in the case of most of the field BDs. For the
young BDs we used the known distances to the clusters/star-forming
regions.  However, the dust properties $\kappa_\nu$ and the dust
temperature $T$ need a more thorough discussion.

The targeted BDs have a large spread in age and we would expect
different evolutionary stages of the disk material if present at all.
Therefore, we will apply two different sets of dust parameters (see,
e.g., \citealt{Henning95}). The ``young'' dust parameters will be
applied to BDs with ages up to 10\,Myr. The ``debris'' dust parameters
will be used for the other objects, which are all older than 100\,Myr.
We assume a constant dust temperature as an approximation to the
temperature distribution of the circumstellar dust.  The next
paragraphs discuss the two sets of dust parameters.

{\sl``Young'' dust:} For dust around young BDs, we choose, for the
sake of comparison, a mass absorption coefficient of
$\kappa_\nu=2\rm\,cm^2g^{-1}$ at 1.3\,mm. The same value of
$\kappa_\nu$ and a gas-to-dust ratio of 100 was applied by
\citet{Beckwith90} to derive disk masses for T\,Tauri disks.  For
the measurements at {850\um}, we assume a wavelength dependence of
$\kappa_\nu\propto\lambda^{-\beta}$ with $\beta=1$, also in
accordance with \cite{Beckwith90}. This leads to
$\kappa_\nu=3\rm\,cm^2g^{-1}$ at 850\um.

The plausible range of dust temperatures is relatively small. We
assume an average temperature of 10 to 20\,K for the dust. This is the
range of the mass-averaged dust temperature in the models for the disk
around CFHT-BD-Tau 4 discussed by \citet{Pascucci03}.

{\sl``Debris'' dust:} The BDs in the Pleiades and the field
objects are presumably older than 100\,Myr.  At
this age one can no longer assume a T\,Tauri-like disk with its
dust properties. If there is a disk left it will resemble debris
disks like the $\beta$\,Pictoris disk. To allow comparisons to
submillimeter observations of debris disks, we adopted the mass
absorption coefficient used by \citet{Dent00}, i.e.
$\kappa_\nu=0.4\ldots1.7\rm\,cm^2g^{-1}$ at {850\um}.  The
gas-to-dust mass ratio in debris disks is controversially discussed
\citep{Thi01,Lecavelier01}. Therefore, we will only give  the dust
masses in these cases.

Since debris disks are optically thin, the circumstellar dust grains
around the old BDs have at least the temperature of interstellar
grains, i.e. 20 to 30\,K depending on the composition and the actual
interstellar radiation field at the position of the BDs. The heating
of the dust by the old BD is negligible, thus, we assume a temperature
between 20 to 30\,K.

Applying the above discussed dust properties to the millimeter
continuum measurements, we obtain the dust masses compiled in Table
\ref{tab:res}. The ranges result from the uncertainties in the mass
absorption coefficient and the temperature.  Furthermore, the
millimeter emission has been measured at two wavelengths for the
detections. The mass estimates derived from the two measurements
differ slightly. For the detections, the masses in Table \ref{tab:res}
correspond to the minimum and maximum mass estimates. In the case of
non-detections, the mass limits are derived from the 3$\sigma$ flux
limits using the dust parameters yielding the highest masses.

The mass of circumstellar dust for CFHT-BD-Tau\,4 is between 1.4 and
7.6\,$M_{\rm E}$ (Earth mass) and for IC\,348\,613 the mass ranges
between 5.4 and 18\,$M_{\rm E}$.  The dust is certainly distributed in
the form of disks as a detailed analyses of CFHT-BD-Tau 4 by
\cite{Pascucci03} show.  The disk masses for the two BDs are
$0.4\ldots2.4\,M_{\rm J}$ (Jupiter mass) and $1.7\ldots5.7\,M_{\rm
  J}$, if we extrapolate the dust masses to disk masses assuming a
gas-to-dust ratio of 100. The upper limits on the circumstellar dust
mass around the other young BDs are around a few Earth masses.

The dust mass constraints for the nearby field BDs are even more
stringent, reaching upper limits of some Moon masses.  Thus, even
around the low-luminosity BDs circumstellar disks cannot be
long-lived. Debris disks around the old BDs, which may be associated
with planetary systems, cannot be ruled out, since we were not
sensitive down to fractions of Moon masses of dust which is the order
of the dust mass around the low-mass main-sequence star $\epsilon$
Eridani \citep{Dent00}.

\subsection{Disks around young BDs}

One important question about BDs and their disks is the
mass ratio of the central objects and the circumstellar material.
The BDs' masses themselves (CFHT-BD-Tau\,4: $M_*=30\ldots75\,M_{\rm J}$
\footnote{\citet{Martin01} derives a mass of $0.03\,M_\odot$ for
  CFHT-BD-Tau 2 and 3 with spectral types M8 and M9, respectively. The
  CFHT-BD-Tau 4 has the spectral type M7 and is more luminous than the 1~Myr
  isochrone. This suggests that CFHT-BD-Tau 4 is more massive, possibly
  close to the stellar/substellar boundary.}%
, \citealt{Martin01}; IC 348 613: $M_*=20\ldots40\,M_{\rm J}$,
\citealt{Preibisch01}) are uncertain as well as the disk masses.
Therefore we do not attempt to estimate their ratio
here. Still, we note that the observations are consistent with a 
disk/BD mass ratio of a few percent as it is found for most 
of the T\,Tauri stars.
 
The detection of some Jupiter masses of matter around BDs is
especially interesting for extrasolar planet searches.
\citet{Guenther03} mention that planets may form around a BD if a
disk of enough mass is present. Their radial velocity survey
targets very low-mass stars and BDs because these targets are
relatively inactive. Furthermore, young planets around BDs would be 
the easiest to detect by direct imaging because the contrast between
the BD and its planet would be much more favorable than for a
planet around a higher-mass star.

\section{Summary}

The observing campaign targeting BDs of several populations
resulted in detections of millimeter emission associated with two
young BDs. The mass estimates for the young BD disks are
0.4\dots2.4\,$M_{\rm
  J}$ and 1.7\dots5.7\,$M_{\rm J}$ for CFHT-BD-Tau\,4 and
IC\,348\,613, respectively. For the other targets, stringent upper
limits on the amount of circumstellar matter were derived from the
measured upper limits on the millimeter continuum flux densities.
To estimate the dust masses, two sets of dust properties had to be
applied: ``Young'' dust properties to BDs younger than 10\,Myr and
``debris'' dust properties to BDs older than 100\,Myr.

The detection of a few Jupiter masses of circumstellar matter
around young BDs is an important result. To ensure this mass
estimate, the dust properties have to be constrained further.
However, a refinement of the dust properties will hardly change
the fact that there are substantial amounts of circumstellar
material around the two BDs CFHT-BD-Tau\,4 and IC\,348\,613. Thus,
the detections make BDs to places of possible planet formation.
This fact opens a new set of targets for extrasolar planet searches,
especially for direct imaging because of the low contrast between
the central object and an prospective planet.

\acknowledgments We thank Remo Tilanus for the valuable support
before, during and after the JCMT observations. The JCMT is operated
by the Joint Astronomy Centre on behalf of the UK Particle Physics and
Astronomy Research Council, the Canadian National Research Council and
the Netherlands Organization for Scientific Research. The observing
run at the JCMT was supported by the Deutsche Forschungs Gemeinschaft
(DFG) through grants Kl 1330/2-1 and Kl 1330/3-1. DA and IP
acknowledge the staff support during the IRAM run 001-02. We thank the
referee for the helpful comments on our work. RK acknowledges support
through the DFG grant He 1935/15-1.


\clearpage
\begin{deluxetable}{rlllrcl}
  \tablewidth{0pt} \tabletypesize{\scriptsize} \tablecaption{Target
    list\label{tab:obs}}
  \tablehead{No.&Target&\multicolumn{2}{c}{Co-ordinates}&
    \colhead{Dist.}&\colhead{Age} &References\\
    &&\colhead{RA(2000)}&\colhead{DEC(2000)}&\colhead{(pc)}&
    \colhead{(Myr)}} \startdata
  &\multicolumn{3}{l}{\bf BDs in Taurus}&140&1&{14}\\
  1\dots&CFHT-BD-Tau 1&$04^h 34^m 15.2^s$&+22\degr50\arcmin31\arcsec\\
  2\dots&CFHT-BD-Tau 2&$04^h 36^m 10.4^s$&+22\degr59\arcmin56\arcsec\\
  3\dots&CFHT-BD-Tau 4&$04^h 39^m 47.3^s$&+26\degr01\arcmin39\arcsec&&&17\\
  &\multicolumn{3}{l}{\bf BD in $\sigma$ Ori cluster}&370&1&{ 2, 15}\\
  4\dots&S Ori 03     &$05^h 39^m 20.8^s$&-02\degr30\arcmin35\arcsec\\
  &\multicolumn{3}{l}{\bf BD in IC\,348}&260&0.5--10&{ 10, 21}\\
  5\dots&IC\,348\,613 &$03^h 44^m 26.9^s$&+32\degr09\arcmin24.8\arcsec\\
  &\multicolumn{3}{l}{\bf BDs in U. Scorpio.}&145&5&{ 1, 18, 22}\\
  6\dots&USco 100     &$16^h 02^m 04.13^s$&-20\degr50\arcmin41.5\arcsec\\
  7\dots&USco 104     &$15^h 57^m 12.66^s$&-23\degr43\arcmin45.3\arcsec\\
  8\dots&USco 112     &$16^h 00^m 26.57^s$&-20\degr56\arcmin32.0\arcsec\\
  9\dots&USco 128     &$15^h 59^m 11.20^s$&-23\degr37\arcmin59.0\arcsec\\
  \hline
  &\multicolumn{3}{l}{\bf BDs in the Pleiades}&116&120&{ 4, 13, 20}\\
  10\dots&NPL 36          &$03^h 48^m 19.1^s$&+24\degr25\arcmin15\arcsec\\
  11\dots&NPL 37          &$03^h 47^m 12.1^s$&+24\degr28\arcmin31\arcsec\\
  12\dots&NPL 38          &$03^h 47^m 50.4^s$&+23\degr54\arcmin49\arcsec\\
  13\dots&NPL 30 (Teide 1)&$03^h 47^m 17.9^s$&+24\degr22\arcmin32\arcsec\\
  &\multicolumn{3}{l}{\bf Field
    BDs}&&\raisebox{-.7ex}{$\stackrel{\displaystyle>}{\sim}$}100\tablenotemark{a}&\\
  14\dots&Gl 229B &$06^h 10^m 35.1^s$&-21\degr51\arcmin18\arcsec&
  5.8&500&{  9, 12, 16}\\
  15\dots&2MASSI J0746425+200032\tablenotemark{b} &$07^h 46^m
  42.5^s$&+20\degr00\arcmin32\arcsec&
  12.3&&{ 3, 7}\\
  16\dots&2MASSI J0825196+211552 &$08^h 25^m
  19.6^s$&+21\degr15\arcmin52\arcsec&
  12.5&&{ 3, 7}\\
  17\dots&LHS 2397a\tablenotemark{c,d} &$11^h 21^m
  49.3^s$&-13\degr13\arcmin09\arcsec&
  14.2&2--12\,Gyr&{ 5, 6, 7, 8}\\
  18\dots&Kelu 1 &$13^h 05^m 40.2^s$&-25\degr41\arcmin06\arcsec&
  19.2&0.3--1\,Gyr&{ 3, 7, 11, 19}\\
  19\dots&TVLM 868-110639\tablenotemark{d} &$15^h 10^m
  17.2^s$&-02\degr41\arcmin07\arcsec&
  25.4&&{ 3, 7, 8}\\
  \hline 
  \enddata 
  
  \tablenotetext{a}{Lower limit for the objects where no individual
    age estimate is available. Comparison with evolutionary tracks
    suggests ages of the order of 100 Myr and higher, see (3).}
  
  \tablenotetext{b}{Binary BD}
  
  \tablenotetext{c}{BD Companion (15)}
  
  \tablenotetext{d}{Objects at the substellar boundary}
  
  \tablerefs{(1) \citealt{Ardila00}; (2) \citealt{Bejar01}; (3)
    \citealt{Dahn02}; (4) \citealt{Festin98a,Festin98b}; (5)
    \citealt{Freed03}; (6) \citealt{Gliese91}; (7)
    \citealt{Kirkpatrick00}; (8) \citealt{Leggett98}; (9)
    \citealt{Leggett99}; (10) \citealt{Luhman99}; (11)
    \citealt{Martin99}; (12) \citealt{Nakajima95}; (13)
    \citealt{Mermilliod97}; (14) \citealt{Martin01}; (15)
    \citealt{Mokler02}; (16) \citealt{Oppenheimer95}; (17)
    \citealt{Pascucci03}; (18) \citealt{Preibisch98}; (19)
    \citealt{Ruiz97}; (20) \citealt{Stauffer98}; (21)
    \citealt{Scholz99}; (22) \citealt{Zeeuw99}}

\end{deluxetable}
\clearpage

\begin{deluxetable}{rccr}
\tablewidth{0pt}
\tablecaption{Observational results\label{tab:res}}
\tablehead{
No.&\colhead{Flux@850\um}&\colhead{Flux@1.3\,mm}&\colhead{Dust~mass}\\
&\colhead{(mJy)}&\colhead{(mJy)}&\colhead{($M_{\rm E}$)}}
\startdata
\multicolumn{4}{l}{\bf BDs in Taurus}\\
1\dots    &&$<$2.73&$<\phn4.8\phn$\\
2\dots    &$<$5.81   &&$<\phn4.1\phn$\\
3\dots    &$10.8\pm1.8$  &$2.1\pm0.6$&\hfill${\bf1.4\ldots7.6}$\\
\multicolumn{4}{l}{\bf BD in $\sigma$ Ori cluster}\\
4\dots    &&$<$2.97&$<32\phd\phn\phn$\\
\multicolumn{4}{l}{\bf BD in IC\,348}\\
5\dots  &$7.6\pm2.4$ &$2.8\pm0.8$&\hfill${\bf5.4\ldots18\phd}$\\
\multicolumn{4}{l}{\bf BDs in Upper Scorpio}\\
6\dots  & $<$4.09   &&$<\phn3.1\phn$\\
7\dots  & $<$5.27   &&$<\phn3.8\phn$\\
8\dots  & $<$6.11   &&$<\phn4.4\phn$\\
9\dots  & $<$8.35   &&$<\phn6.3\phn$\\
\hline\hline
\multicolumn{4}{l}{\bf BDs in the Pleiades}\\
10\dots      & $<$6.04   &&$<\phn6.7\phn$\\
11\dots      & $<$4.18   &&$<\phn4.6\phn$\\
12\dots      & $<$4.04   &&$<\phn4.4\phn$\\
13\dots      & $<$4.39   &&$<\phn4.9\phn$\\
\multicolumn{4}{l}{\bf Field BDs}\\
14\dots      & $<$9.41   &&$<\phn0.03$\\
15\dots      & $<$3.76   &&$<\phn0.05$\\
16\dots      & $<$5.16   &&$<\phn0.07$\\
17\dots     &  $<$4.79   &&$<\phn0.08$\\
18\dots      & $<$2.15   &&$<\phn0.06$\\
19\dots      & $<$6.73   &&$<\phn0.36$\\
\hline
\enddata
\end{deluxetable}
\end{document}